\documentclass[runningheads]{llncs}
\usepackage[T1]{fontenc}
\usepackage{graphicx}
\usepackage{bbding}
\usepackage{graphicx}
\usepackage{amsmath}
\usepackage{orcidlink}
\usepackage{booktabs}
\usepackage{tabularx}
\usepackage{makecell}
\usepackage{adjustbox}
\hypersetup{hidelinks}

\begin{document}
\title{What is the focus of XAI in UI design? Prioritizing UI design principles for enhancing XAI user experience}
\titlerunning{Prioritizing UI Design Principles in XAI for User Experience}
%
\author{Dian Lei\orcidlink{0009-0002-5585-5930} \and
Yao He \and
Jianyou Zeng\inst{(}\Envelope\inst{)}}
\authorrunning{D. Lei et al.}
%
\institute{China University of Geosciences, Wuhan, 430000, China \\
\email{\{thunder98, heyao, jianyou\}@cug.edu.cn}}

\maketitle              
\begin{abstract}
With the widespread application of artificial intelligence (AI), the explainable AI (XAI) field has undergone a notable resurgence. In this background, the importance of user experience in XAI has become increasingly prominent. Simultaneously, the user interface (UI) serves as a crucial link between XAI and users. However, despite the existence of UI design principles for XAI, there is a lack of prioritization based on their significance. This will lead practitioners to have a vague understanding of different design principles, making it difficult to allocate design space reasonably and emphasize design focal points. This paper aims to prioritize four design principles, providing clear guidance for UI design in XAI. Initially, we conducted a lightweight summary to derive five user experience standards for non-expert users in XAI. Subsequently, we developed four corresponding webpage prototypes for the four design principles. Nineteen participants then interacted with these prototypes, providing ratings based on five user experience standards, and We calculated the weights of the design principles. Our findings indicate that, for non-expert users, "sensitivity" is the optimal UI design principle (weight = 0.3296), followed by "flexibility" (weight = 0.3014). Finally, we engage in further discussion and summarization of our research results, and present future works and limitations.

\end{abstract}
\section{Introduction}

AI has permeated every facet of our lives and gradually integrated into our daily routines. The widespread popularity of large language models (LLMs) has further intensified AI's impact on our daily lives. However, the explanation of AI output is not only a requirement for user experience but also a legal mandate for the implementation of AI, such as the European Union's GDPR~\cite{regulation2016regulation}. Consequently, the field of XAI has entered its third wave of research, with numerous emerging XAI technologies. In the early stages of XAI research, there was a lack of user involvement, relying primarily on the preferences of technical experts. The opinions of the Human-Computer Interaction (HCI) community were often overlooked or even rejected~\cite{Yang_Scuito_Zimmerman_Forlizzi_Steinfeld_2018}. This has resulted in a significant focus on algorithms and a disconnect from the actual usage environment of XAI. Later on, many HCI researchers recognized the importance of a user-centric perspective and attempted to shift the focus of XAI research from algorithms to the human~\cite{Liao_Varshney_2022,Ehsan_Riedl_2020,Miller_2019,Mueller_Hoffman_Clancey_Emrey_Klein_2019}.The UI design for XAI has also garnered more attention, with UI being considered a crucial pathway for XAI output. Many researchers view UI as the second step in the entire XAI application process, serving as the bridge for presenting humanized outputs from specialized XAI data results. To reduce unnecessary text, we will generalize the UI designed for XAI as XUI, defined as "the sum of outputs of an XAI system that the user can directly interact with."~\cite{Chromik_Butz_2021}

However, despite numerous attempts to enhance user experience in XAI, the current state of affairs still reflects a disconnect between user needs and existing XAI systems~\cite{Liao_Varshney_2022,Miller_2019}. Moreover, there is limited research on how humans perceive XAI and their expectations of XAI systems~\cite{Vainio-Pekka_Agbese_Jantunen_Vakkuri_Mikkonen_Rousi_Abrahamsson_2023}. Thus, improving the user experience with a human-centered approach remains a worthwhile direction in XAI. There is existing research that has summarized XUI design principles~\cite{Chromik_Butz_2021}, but it has not prioritized  weights to these principles. The design space for XUI is limited, and excessive content may lead to cognitive overload and even psychological conflicts~\cite{Tsai_Brusilovsky_2019}. Therefore, this vague understanding of design principles will lead to a lack of focal points of design and an inability to reasonably allocate design space. Lastly, evaluations of XAI often neglect user experience assessments~\cite{ali2023explainable,van2021evaluating}. While some research exists on XAI user experience evaluations, many standards are tailored for domain-specific professionals, creating a mismatch for non-expert users. Details are further summarized in sect.~\ref{sec3}.

To address these issues, we conducted a quantitative experiment aimed at prioritizing the XUI design principles that enhance user experience. Initially, we developed four webpage prototypes corresponding to four design principles. Then, Users rated these prototypes using the Analytic Hierarchy Process (AHP) to determine their weights~\cite{saaty1988analytic}, respectively. Additionally, we conducted qualitative interviews with users after they completed the quantitative experiment to validate the conclusions and address potential shortcomings in the research.

Through quantitative analysis of the data,  we found that among the five XAI user experience standards, trust and understandability are the most important, with weights of 0.2903 and 0.2398, respectively. Sensitivity and flexibility are identified as the most critical XUI design principles, with weights of 0.3296 and 0.3014. We also obtained the weights of four XUI design principles under the five XAI user experience standards. The contribution of this article is twofold: 

1. We provided weighted priorities for design principles aimed at enhancing  XUI user experience, offering clear guidance for practitioners to allocate XUI design space reasonably.

2. Taking a Human-Centered XAI (HCXAI) perspective, we offered a lightweight summary of user experience standards for non-expert users in XAI. This provides subsequent researchers with a reference framework for better understanding and meeting the expectations of non-expert users in XAI.

\section{Related Work}
In this section, we first review the current status and shortcomings of user experience in the XAI field and then explore the research content related to XUI.

\subsection{\textbf{User Experience in XAI}}
   Research on XAI has a long history, the first generation of XAI systems began to appear in the late 1970s. However, contemporary XAI systems still face challenges from both the first and second generations, particularly in lacking user experience ~\cite{Mueller_Hoffman_Clancey_Emrey_Klein_2019}. In recent years, many HCI researchers have endeavored to address this issue. For example, Springer and Whittaker enhanced the transparency and user experience of intelligent systems through progressive disclosure~\cite{Springer_Whittaker_2018}. Ferreira and Monteiro, in their literature review, observed a general lack of focus on user experience in XAI research outside the HCI community and emphasized the importance of user experience~\cite{Ferreira_Monteiro_2020}. Ehsan and Riedl proposed an approach that places humans at the center of XAI, known as HCXAI. Liao et al~\cite{Ehsan_Riedl_2020}. Liao et al. developed an XAI question bank to meet user understanding needs~\cite{Liao_Gruen_Miller_2020}. However, related studies point out two major issues with the user experience in XAI. First, as highlighted in the papers by Liao and miller~\cite{Liao_Varshney_2022,Miller_2019}, the existing XAI systems still suffer from a disconnect with user requirements, leading to the "inmates running the asylum" problem. Second, evaluations of XAI primarily focus on interpretability (model performance), with user evaluations often being overlooked~\cite{van2021evaluating,ali2023explainable}.

\subsection{\textbf{UI Design for XAI}}
  UI is crucial for XAI, serving as the bridge between users and XAI systems. Program such as DARPA's XAI and the study by Danilevsky et al. roughly divide the XAI process into two stages: the generation of raw explanations by interpretable models, followed by translation through UI into understandable content for the general public\cite{danilevsky2020survey,Gunning_Aha_2019}. Therefore, many researchers have made efforts in UI design for XAI. For instance, Hohman et al. designed Gamut, an interactive explainable interface targeting expert users~\cite{Hohman_Head_Caruana_DeLine_Drucker_2019}. Rjoob developed a user interface for XAI generating Automated ECG (Electrocardiology) Interpretations~\cite{rjoob2021towards}. Janet and Hani designed XAI interfaces tailored for finance professionals~\cite{Adams_Hagras_2020}. Hao-Fei Cheng and collaborators designed various explainable interfaces, including interactive and white-box, for an AI system used in university admissions~\cite{Cheng_Wang_Zhang_Oâ€™Connell_Gray_Harper_Zhu_2019}. Liao, adopting a scenario-based design approach, created a UI design aiming for social transparency in AI systems~\cite{Liao_Varshney_2022}. It is noticeable that existing XUI designs have relatively limited focus on non-expert users. This may be attributed to XAI historically catering to expert users in various domains. However, with the popularity of LLMs, such as ChatGPT, XAI stakeholders and application scenarios are rapid growth~\cite{Liao_Vaughan_2023}. The importance of XUI for ordinary non-expert users continues to increase. 
\section{XAI User Experience Standards for Non-Expert Users}
\label{sec3}
After analyzing multiple literature on XAI user experience standards, this study provides a lightweight summary of the composition of user experience standards(see Table~\ref{tab1}). We identified some shortcomings in existing XAI user experience standards. Firstly, there is currently no complete consensus on user experience standards for XAI, and there are too many standards related to XAI user experience, causing difficulty in flexible application during the evaluation process. Secondly, existing XAI user experience standards lack a clear definition of their target audience. Therefore, there are many standards that are not applicable to non-expert users and that non-expert users do not care about in practical use, such as Parsimony, Causality, Correct rate, etc.

\begin{table}[h]
  \centering
  \renewcommand{\arraystretch}{1.4}
  \caption{Summary of XAI user experience standards}
  \label{tab1}
  \resizebox{\textwidth}{!}{%
    \begin{tabular}{l|l|m{0.7\linewidth}}
      \hline
      \textbf{No.} & \textbf{Author(s)} & \textbf{XAI user experience standards} \\
      \hline
      01 & Sajid et al.~\cite{ali2023explainable}. & Understandability; Satisfaction; Trust; Transparency; Explanation; Trust \\
      \hline
      02 & Samuli et al.~\cite{laato2022explain} & Intelligibility, Comprehensibility, Interpretability; Trust; Transparency; Controllability \\
      \hline
      03 & Markus et al.~\cite{langer2021we} & Understandable; Satisfaction; Explanation \\
      \hline
      04 & Jasper et al.~\cite{van2021evaluating} & Understandable; Persuasion; Correct rate; Accuracy rate \\
      \hline
      05 & Juliana J \& Mateus~\cite{Ferreira_Monteiro_2020} & Adoption rate; Acceptance; Satisfaction; Engagement; Persuasion; Continued use \\
      \hline
      06 & Sule et al.~\cite{Anjomshoae_Najjar_Calvaresi_FrÃ¤mling_2019} & Usefulness; Naturalness; Trust; Transparency; Controllability \\
      \hline
      07 & Martijn et al.~\cite{Millecamp_Htun_Conati_Verbert_2019} & Effectiveness; Understandability; Trust; Novelty; Satisfaction; Confidence \\
      \hline
      08 & Markus et al.~\cite{Mohseni_Zarei_Ragan_2021} & Trust; Explanation; Satisfaction \\
      \hline
      09 & Robert et al.~\cite{Hoffman_Mueller_Klein_Litman_2019} & Explanation; Satisfaction; Understandability; Curiosity; Trust \\
      \hline
      10 & Nava~\cite{Tintarev_2007} & Transparency; Scrutability; Trustworthiness; Effectiveness; Persuasiveness; Efficiency; Satisfaction \\
      \hline
      11 & Tim~\cite{Miller_2019} & Coherence; Simplicity; Generality; Truth; Explanation \\
      \hline
      12 & Aniek~\cite{markus2021role} & Clarity; Parsimony; Completeness; Soundness \\
      \hline
      13 & David \& David~\cite{Gunning_Aha_2019} & Satisfaction; Trust; Predictability; Understandable; Correct rate \\
      \hline
      14 & Nadia \& Marco~\cite{burkart2021survey} & Trust; Transferability; Causality; Informativeness; Accountability; Transparency \\
      \hline
      15 & Shane et al.~\cite{Mueller_Hoffman_Clancey_Emrey_Klein_2019} & Explanation; Trust; Reliance; Predictability \\
      \hline
    \end{tabular}
  }
\end{table}

To address these issues, this study adopts the HCXAI perspective to filter out standards that do not meet the needs of non-expert users. In other words, the focus is on standards that truly reflect the user experience for non-expert users, excluding any standards irrelevant to their experience. Specifically, three different levels are used to integrate XAI standards for non-expert user experience, resulting in five indicators that are truly applicable to non-expert users, the details see Table~\ref{tab2}. The specific summarized information is as follows:

\subsection{Universal User Experience Level : Satisfaction} Universal user experience standards are prerequisites for any system aiming for a good user experience, similar to constituting the "baseline" for a good user experience. This study uses "satisfaction" to encompass universal user experience indicators. Satisfaction can comprehensively reflect the system's usability and the user's psychological pleasure, making it a metric for measuring the overall experience of non-expert users.
\subsection{Excellent Explanation Tool Level : Persuasiveness, Efficiency}  Explanation is a crucial component of XAI, and the effectiveness of explanations directly influences the user experience. Therefore, an XAI system for non-expert users should meet the requirements of an excellent explanation tool. Some standards for excellent explanation tools overlap with unique XAI user experience standards, which we will not repeat. In this study, we choose"persuasiveness" and "efficiency" as the criteria for excellent explanation tools. Persuasiveness is a key factor in the effectiveness of an XAI system, and good persuasiveness not only enhances the user experience but can also influence user behavior for better decision-making~\cite{Dragoni_Donadello_Eccher_2020}. On the other hand, efficiency is crucial for user experience, providing users with a sense of fluency and confidence~\cite{confalonieri2021using}. For non-expert users interacting with XAI systems, because the XAI systems they use lean towards frequent application, a smooth user experience is highly essential.
\subsection{Unique XAI User Experience Level : Understandability, Trust} XAI systems differ from ordinary products, and users have higher expectations for attributes such as transparency, trust, and reliability. Establishing unique experiences for XAI users contributes to a more in-depth evaluation of XAI user experience. This study uses "understandability" and "trust" to reflect these unique standards. Understandability has long been a persistent issue in XAI. For example, many XAI algorithms generate graphical results, such as LIME and SHAP, which can be challenging for non-expert users to understand~\cite{kaur2020interpreting,Xu_Dainoff_Ge_Gao_2023}. Additionally, the degree of understanding of explanations is higher when they align with the user's mental model~\cite{Xie_Gao_Chen_2019}. Therefore, understandability can reflect the degree of matching between the XAI system and the user's mental model. In systems involving risks, the level of trust that users have in the system directly determines their experience~\cite{Cahour_Forzy_2009}. Trust is one of the most important user experience characteristics for XAI aimed at ordinary non-expert users. Existing studies suggest a high dependence between trust, transparency, and controllability~\cite{Ehsan_Riedl_2020,Liao_Vaughan_2023}. And, Research suggests that trust in intelligent systems stems from control and transparency~\cite{Cahour_Forzy_2009}. So, Trust as an indicator can effectively reflect the non-expert user's experience with the controllability and transparency of the XAI system. 

\begin{table}[h]
  \centering
  \renewcommand{\arraystretch}{2}
  \caption{XAI User Experience Standards}
  \label{tab2}
  \resizebox{\textwidth}{!}{%
    \begin{tabular}{l|m{0.7\linewidth}}
      \hline
      \textbf{Standard} & \textbf{Description} \\
      \hline
      Satisfaction & This standard measures whether users gain satisfaction during use. \\
      \hline
      Trust & The standard of trust involves whether users increase their trust in the AI system because of the explanation method. \\
      \hline
      Persuasiveness & The persuasiveness standard focuses on whether users feel that XAI’s explanation is convincing. \\
      \hline
      Efficiency & The efficiency standard refers to whether users feel that they have gained higher speed when understanding XAI. \\
      \hline
      Understandability & The standard of understandability examines whether the content of XAI is easy for users to understand. \\
      \hline
    \end{tabular}
  }
\end{table}

\section{Method}
In order to explore the weight of design principles in XUI, we employed a mixed-method approach for experimentation and data processing. Firstly, we created four web prototypes based on four XUI design principles, and we used each of the four XUI design principles to explain the same AI medical conclusion, in this study, we assume that the user is diagnosed with coronary heart disease and has corresponding symptoms and abnormal physiological indicators. Secondly, we used the Analytic Hierarchy Process (AHP) method for quantitative analysis of user experiences with the four web prototypes~\cite{saaty1988analytic}, obtaining specific weight information. Finally, after the experiment, We conducted qualitative interviews with participants to validate the conclusions drawn from our previous quantitative analysis and to supplement areas that might have been overlooked during the experimental process.
\subsection{Design principles for Enhancing XUI User Experience}
In the context of XUI design principles, we primarily adopted the principles proposed by Chromik and Butz   ~\cite{Chromik_Butz_2021} in their SLR article. However, this paper introduced some modifications to the aspect of naturalness to ensure its distinctiveness from the other three design principles. For specific design principles and explanations, see Table~\ref{tab3}.

\begin{table}[h]
    \centering
    \caption{Design principles for XUI}
    \label{tab3}
    \renewcommand{\arraystretch}{2}
    \resizebox{\textwidth}{!}{%
        \begin{tabular}{c|m{0.7\linewidth}}
            \hline
            \textbf{Design principle} & \textbf{Description} \\
            \hline
            A: Naturalness & This principle aims to enhance the logic and accuracy of explanations through natural language. It achieves this by using the substantial information content and rapid rationalization characteristic of natural language to generate detailed and logically sound explanations. \\
            \hline
            B: Responsiveness & The principle of responsiveness aims to dynamically respond to the initial interpretation according to the user's needs, mainly through progressive disclosures of information to meet the user's needs. This method not only helps reduce the cognitive load of users but also satisfies users with different depths of understanding. \\
            \hline
            C: Flexibility & The principle of flexibility encourages the use of multiple different ways of explanation to form a triangular and mutually supporting explanation mechanism and enhance the comprehensiveness and credibility of explanations. \\
            \hline
            D: Sensitivity & The principle of sensitivity emphasizes the continuous adjustment of explanation principles according to the user's psychological state and usage scenarios to ensure the adaptability and effectiveness of explanations. \\
            \hline
        \end{tabular}
    }
\end{table}

\subsection{Prototype Design}
We constructed a fictitious online health assessment scenario. because, in the context of AI inferences related to health matters, users have a stronger demand for explanations~\cite{Holzinger_Biemann_Pattichis_Kell_2017}. This helps capture the attention of our participants. Our primary objective is provide an environment to experience various XUI design principles and to gather feedback data in subsequent evaluations.

In the design practice. firstly, we used feature-based explanation style. Secondly, we designed the UI in the form of conversational agents. Finally, all our explanatory content is in the form of post hoc local explanation. This is mainly due to the following reasons: 1) Previous research indicates that the feature-based explanation style performs well in Online Symptom Checkers (OSCs), sharing similarities with the experimental design of this study~\cite{Tsai_You_Gui_Kou_Carroll_2021}. 2) The natural human demand for social explanations leads us to prefer conversational styles of explanation, and conversational methods are considered one of the most promising approaches in intelligent system explanations~\cite{Miller_2019}. And intelligent agents can be easily embedded into various systems as tools for explanation~\cite{Mueller_Hoffman_Clancey_Emrey_Klein_2019}. Additionally, popular LLMs provide extensive technical support for conversational agents. 3) Research shows that users prefer Local Explanation in practical usage~\cite{Radensky_Downey_Lo_Popovic_Weld_2022}. At the same time, XAI technologies for local explanations are also richer. The specific details of XUI are as follows:
\subsubsection{Natureness}
\begin{figure}
\includegraphics[width=\textwidth]{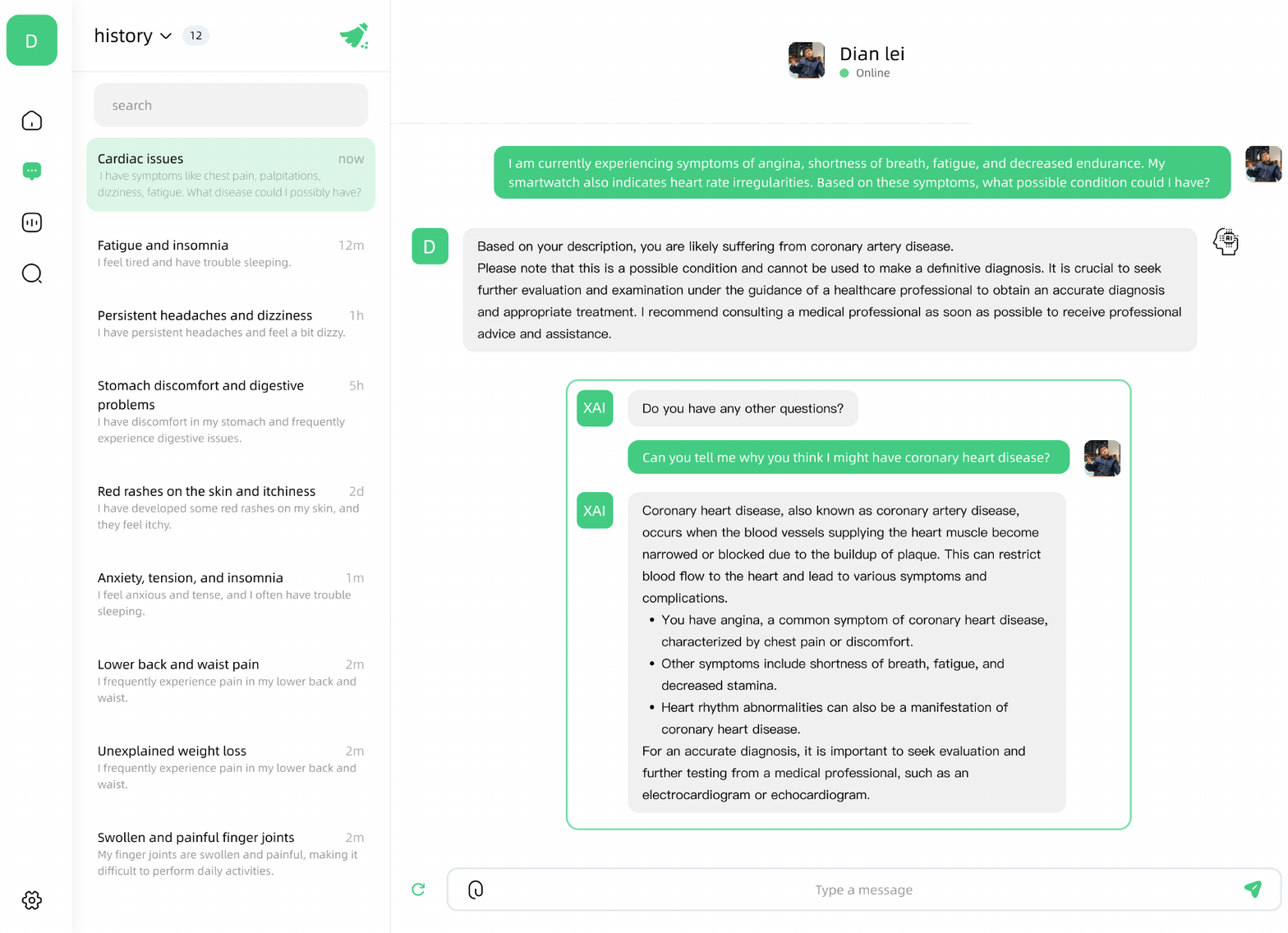}
\caption{The XUI of Natureness.} \label{fig1}
\end{figure}

Although this may seem like a very common explanatory approach, its explanations are not only rich in information content but also quite accurate. This gives it a certain advantage in systems involving risks. For example, research indicates that when users become aware of their health anomalies, they prefer comprehensive and accurate explanations~\cite{Tsai_You_Gui_Kou_Carroll_2021}. Additionally, different cultural backgrounds and preferences may lead users to prefer textual explanations~\cite{klein2014influencing}. Furthermore, it also offers advantages in terms of faster generation speed and rationalization speed~\cite{Ehsan_Tambwekar_Chan_Harrison_Riedl_2019}. See the specific design in Fig.~\ref{fig1}.

\subsubsection{Responsiveness}
\begin{figure}
\includegraphics[width=\textwidth]{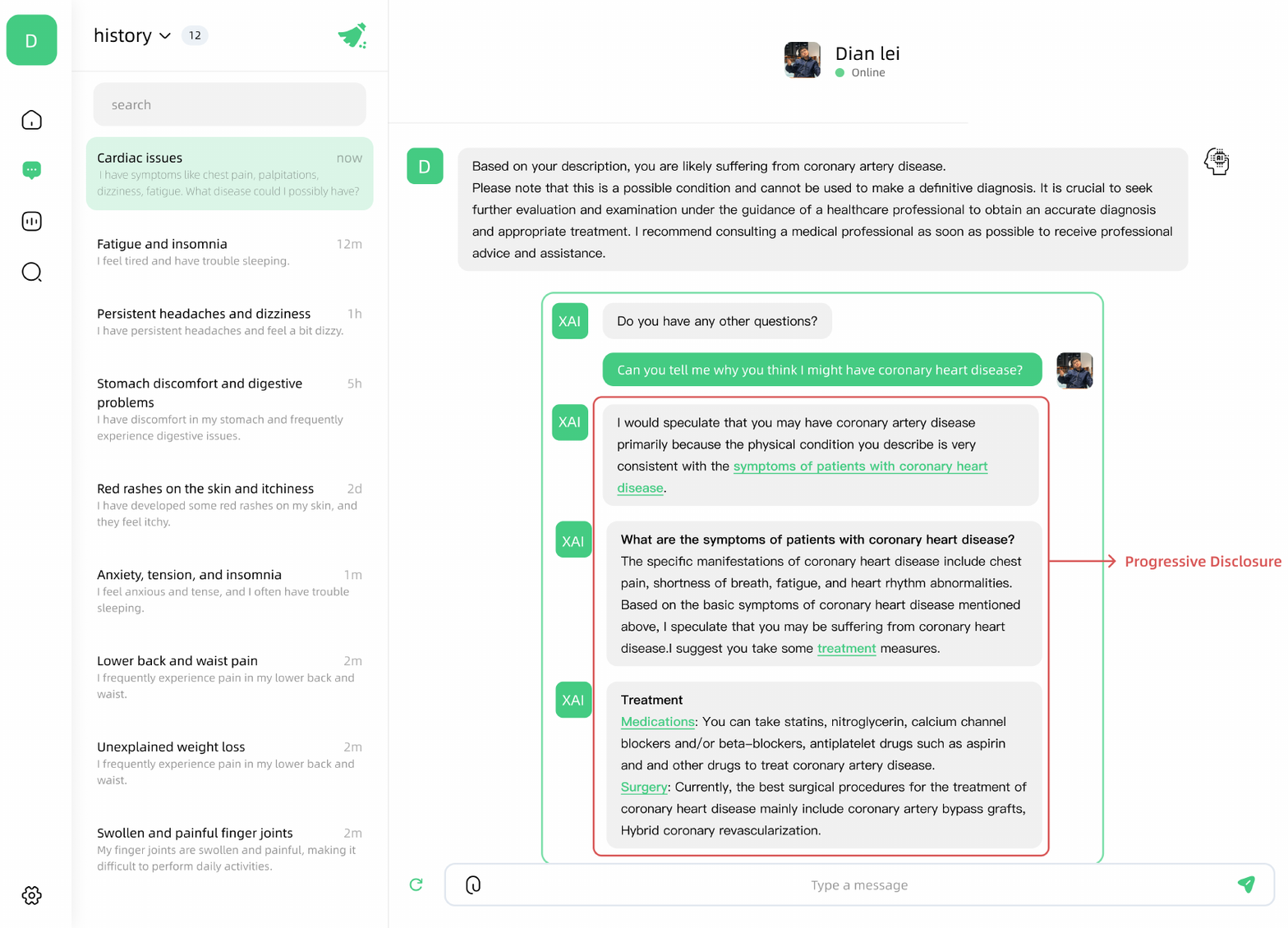}
\caption{The XUI of Responsiveness.} \label{fig2}
\end{figure}

We created an interactive XUI through Progressive Disclosure to meet users' responsiveness needs. Non-expert users dislike explanations that require much effort~\cite{gregor1999explanations}, and this approach can significantly reduce the likelihood of user information overload. By progressively providing information, it becomes easier for users to obtain personalized depth of explanation. Specifically, we initially provide users with a brief natural language explanation of why they are diagnosed with coronary heart disease. Next, users are free to choose additional information they want to explore further, such as an introduction to coronary heart disease or its symptoms. Finally, users can delve into how to treat the disease. We limited the levels of Progressive Disclosure to two layers because exceeding two layers can cause users to lose their way in the hierarchy~\cite{Experience}. See the specific design in Fig.~\ref{fig2}.

\subsubsection{Flexibility}
\begin{figure}
\includegraphics[width=\textwidth]{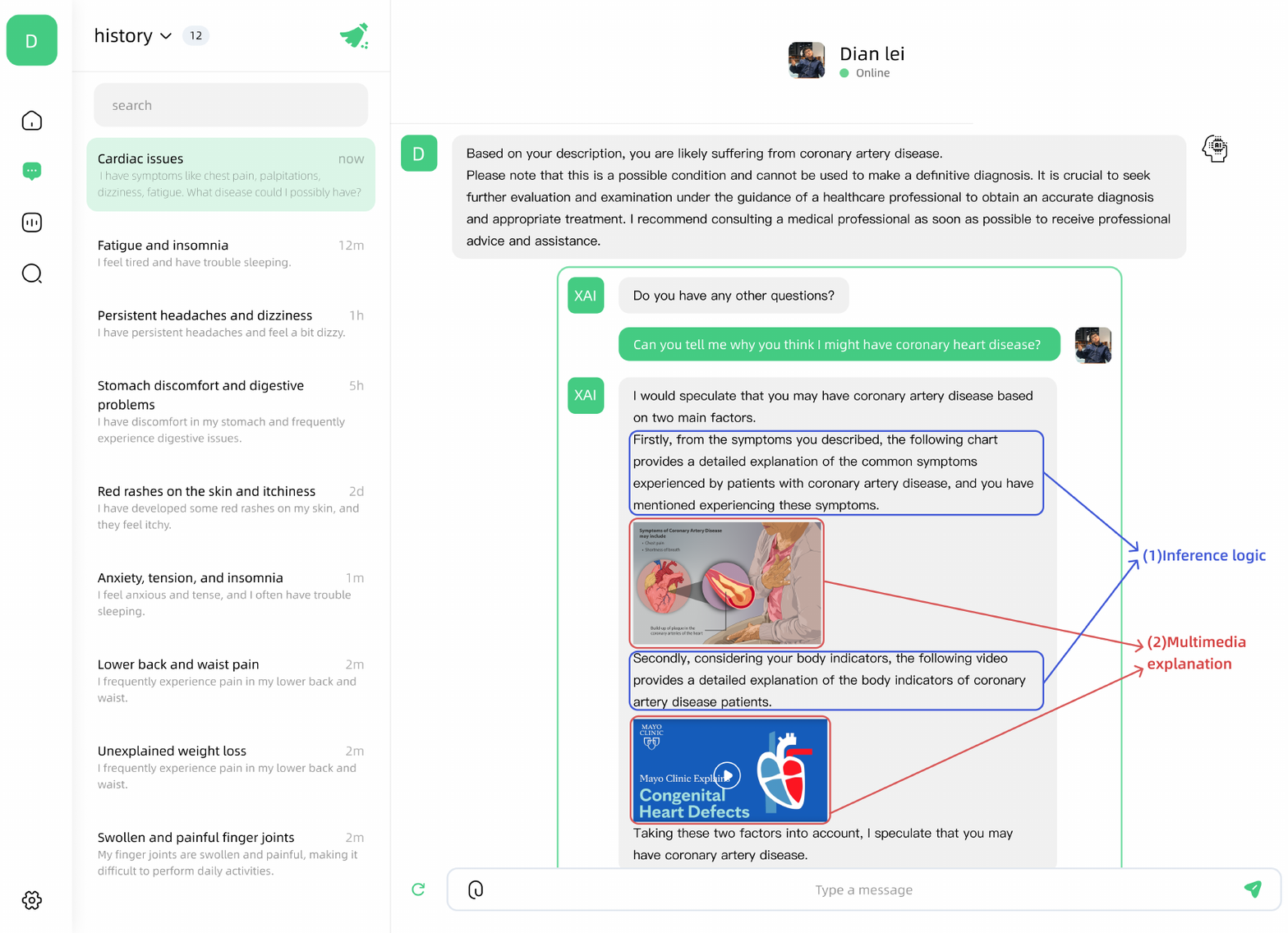}
\caption{The XUI of Flexibility.}
\label{fig3}
\end{figure}

Humans seek understanding through diverse ways~\cite{Chromik_Butz_2021}. Similar to our research, we frequently use triangulation to reduce errors. Diverse explanatory approaches play a positive role when users are suspicious of the results. In the flexible XUI design, we emphasize corroborating various forms of explanatory materials and logical explanatory methods. In our XUI, we have set up two different diagnostic explanations for coronary heart disease: 1) Inference logic: a) inferring based on user self-reported symptoms; b) inferring based on user self-reported physiological indicators. 2) Multimedia explanation: providing detailed explanations of users' symptoms and their self-described correspondence through videos and images\footnote{https://www.youtube.com/watch?v=x6VrwrIonc0}\footnote{https://www.myupchar.com/en/disease/coronary-artery-disease}. See the specific design in Fig.~\ref{fig3}.

\subsubsection{Sensitivity}
\begin{figure}
\includegraphics[width=\textwidth]{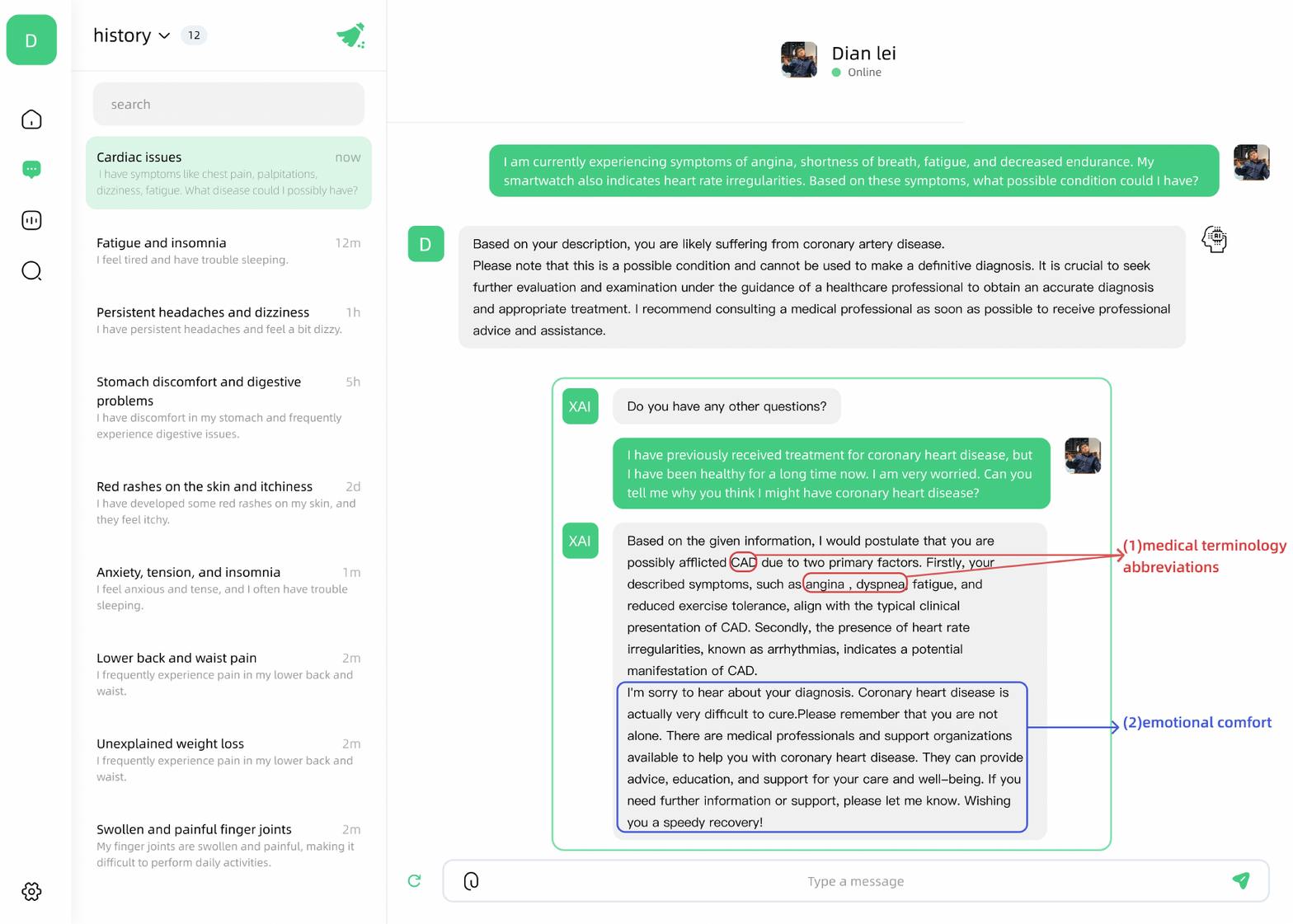}
\caption{The XUI of Sensitivity.} \label{fig4}
\end{figure}

This principle is primarily designed to address the diverse explanation needs of users. Therefore, it requires XUI to keenly grasp changes in user explanation needs and dynamically generate corresponding explanations based on the user's psychological model and state in real-time. To show the characteristics of sensitivity, we introduce a new user context. We assume that the user had previously suffered from coronary heart disease but has been healthy for a long time. However, the AI re-diagnosed them with coronary heart disease. XAI adjusts its responses based on this new user background to demonstrate the system's adaptive adjustment to the user's background and psychological state. For instance, in this XUI, there are two instances: 1) When the AI recognizes that the user has a basic understanding of medical knowledge and treatment methods, the AI begins to attempt direct communication with the user using medical terminology abbreviations. 2) Considering the user's anxious mindset upon learning about the recurrence, the system provides emotional comfort and suggests ways to alleviate the disease. See the specific design in Fig.~\ref{fig4}.

\subsection{Participants}

To finish the experiment, we recruited 19 adult participants, including both teachers and students. The age range of the participants was 20 to 52 years (M = 28.65, SD = 9.59), comprising 9 females and 10 males. We deliberately selected individuals with diverse professional backgrounds to comprehensively assess the effectiveness of the XAI system across different demographics. The participants represented various age groups and genders to ensure the broad applicability of the experimental results. All participants possessed an adequate level of cultural literacy, the necessary knowledge, and the skills to comprehend the information presented by the XAI system. Moreover, all participants had no experience in using XAI systems. Before the start of the experiment, we provided detailed explanations to the participants to ensure their understanding of the XAI system's features, the experiment's objectives, and the meaning of the AHP scale. All participants volunteered to take part in the study, and each received a gift of approximately \$10 after completion.

\subsection{Data Collection and Analysis}

Participants provided evaluations for each principle according to the AHP scoring table (see Table~\ref{tab4}). Two rating tables were excluded due to the failure of the consistency check.

\begin{table}[h]
    \centering
    \label{tab4}
    \renewcommand{\arraystretch}{1.3}
    \caption{AHP assessment ratio scale and description}
    \begin{tabular}{c|m{0.25\linewidth}|m{0.5\linewidth}}
        \hline
        \textbf{Scale} & \textbf{Definition} & \textbf{Explanation} \\
        \hline
        1 & As important as & Means $i$ factors are as important as $j$ factors \\
        3 & Slightly more important & Means $i$ factors are slightly more important than $j$ factors \\
        5 & Obviously more important & Means $i$ factor is obviously more important than $j$ factor \\
        7 & More important & Means $i$ factor is more important than $j$ factor \\
        9 & Extremely important & Means $i$ factor is extremely important than $j$ factor \\
        2, 4, 6, 8 & Median & The median value of the two adjacent judgments \\
        Count backwards & Relative count backwards & When the $j$ factor is compared with the $i$ factor, the judgment value is $a_{ij} = \frac{1}{a_{ji}}$ \\
        \hline
    \end{tabular}
\end{table}

\begin{figure}
\includegraphics[width=\textwidth]{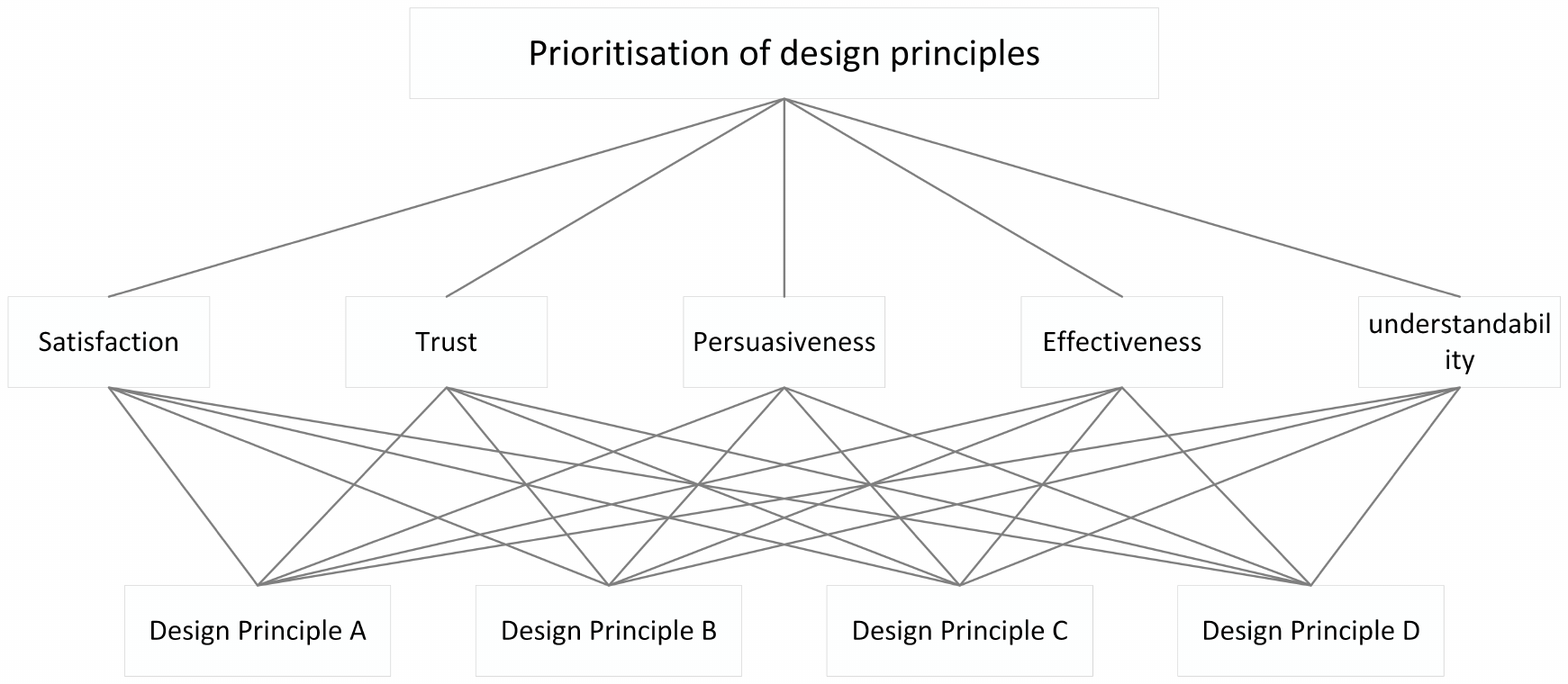}
\caption{The AHP model} \label{fig5}
\end{figure}

Subsequently, we proceeded with model construction. Initially, we used the design principles of the four XUIs to form the decision layer of the AHP model. Following that, we used five XAI experience standards tailored for ordinary users to constitute the criteria layer, as depicted in Fig.~\ref{fig5}.

The following illustrates the process of determining the weights of the 5 XAI user experience standards using the AHP method, using the example of a participant (P1). Matrix processing mainly involves the following steps:

1. We constructed the corresponding judgment matrix A based on user ratings, as shown in Formula~\ref{eq:matrix1} (Matrix diagram in Formula~\ref{eq:matrix2}) :
\begin{equation}
\label{eq:matrix1}
A=  
\renewcommand{\arraystretch}{1.5} 
\setlength{\arraycolsep}{5pt} 
\begin{bmatrix}
    1 & \frac{1}{5} & 3 & 1 & \frac{1}{3} \\
    5 & 1 & 4 & 3 & 3 \\
    \frac{1}{3} & \frac{1}{4} & 1 & \frac{1}{3} & \frac{1}{3} \\
    1 & \frac{1}{3} & 3 & 1 & \frac{1}{5} \\
    3 & \frac{1}{3} & 3 & 5 & 1
\end{bmatrix}
\end{equation}

\begin{equation}
\label{eq:matrix2}
A_{m\times n} =
\begin{bmatrix}
    a_{11} & a_{12} & \cdots & a_{1n} \\
    a_{21} & a_{22} & \cdots & a_{2n} \\
    \vdots & \vdots & \ddots & \vdots \\
    a_{m1} & a_{m2} & \cdots & a_{mn}
\end{bmatrix}
= \left [ a_{ij}\right ]
\end{equation}

2.We used the square root method to obtain its column vector, i.e., using Formula~\ref{eq:formula1}. Then, we normalized it using Formula~\ref{eq:formula2}. Consequently, we obtained specific information about the weights of the 5 XAI user experience standards for P1, as shown in Table~\ref{tab5} :

\begin{equation}
\label{eq:formula1}
\bar{\omega_{\tiny i}} = \sqrt[m]{\prod_{j=1}^{m} a_{\tiny ij}}
\end{equation}

\begin{equation}
\label{eq:formula2}
\omega_{\tiny i} = \frac{\bar{\omega}_{\tiny i}}{{\textstyle \sum_{j=1}^{m}\bar{\omega}_{\tiny j}}}
\end{equation}

\begin{table}[h]
\centering
\caption{XAI User Experience Weight for P1}
\renewcommand{\arraystretch}{1.4}
  \begin{tabular}{c|c|c|c|c}
    \hline
    Satisfaction & Trust & Persuasiveness & Efficiency & Understandability \\
    \hline
    0.1083 & 0.4414 & 0.0623 & 0.1123 & 0.2757 \\
    \hline
  \end{tabular}
  \label{tab5}
\end{table}

3.Calculate the largest eigenvalue of the matrix using Formula~\ref{eq:formula3}, and the calculated value is $\lambda$max = 5.4276:

\begin{equation}
\label{eq:formula3}
\lambda_{\tiny \text{max}} = \sum_{i=1}^{n} \frac{(\mathrm{A} \omega)_{\tiny i}}{n \omega_{\tiny i}}
\end{equation}

4. The consistency of the matrix was examined through Formulas~\ref{eq:formula4} and ~\ref{eq:formula5}, and the R.I. value is only related to the order of the judgment matrix, and it is 1.12 in this case. The obtained C.R. value of 0.0955 < 0.1 confirms that it passed the consistency test:

\begin{equation}
\label{eq:formula4}
C.I. = \frac{\lambda_{\tiny \text{max}} - n}{n-1}
\end{equation}

\begin{equation}
\label{eq:formula5}
C.R. = \frac{C.I.}{R.I.}
\end{equation}

5. Repeat this process to explore users' weights for each UI design principle under each user experience criterion. Multiply the weights of the corresponding design principle by the weights obtained for the respective experience criterion, and then sum them up to obtain the total weight of the decision layer.

By repeating these steps for each participant, we obtained specific scores for the four XUI design principles, their preferences for the 5 user experience standards, and the scores of the 4 XUI design principles under different user experience standards.

\subsection{Interviews}

To validate the rationale of our experiments, we decided to conduct interviews with users after the conclusion of the experiments. Through this interview process, we aim to ensure the credibility and effectiveness of the experiments, while also gaining insights into users' subjective experiences and feedback to better comprehend the experimental data comprehensively. The interview questions are as follows:

Q1: Please describe which specific XUI design principles had a significant impact on your user experience during the interaction with the XAI system, and explain the specific ways in which it influenced your experience.

Q2: For each of the five XAI user experience criteria, please discuss which XUI design principles achieved better results.

Q3: In your opinion, in which aspects of UI design further research or improvement is needed to achieve enhanced user interaction and interpretability?

\section{Result}

We obtained various data results through calculations. Specifically, we acquired the weights of users for five XAI user experience standards, shown in Table~\ref{tab6}. Additionally, we obtained the weights for different design principles of XUI, shown in Table~\ref{tab7}. Furthermore, the weights of different design principles of XUI under the five XAI user experience standards are shown in Table~\ref{tab8}. Excluding the data result, and combining the content from interviews, we primarily derived the following results :

\textbf{1. Trust is the most crucial aspect among the XAI user experience standards,} with a weight of 0.2903, followed by Understandability, the second most important standard, with a weight of 0.2398. Responses to Q1 during the interviews also confirmed this observation. Users often consider trust as the foundation for a good XAI user experience. For example, P7 mentioned, "If the system cannot provide enough trust, I find it challenging to have a positive perception of the system. Even if other aspects are well-executed, I am likely to maintain a skeptical outlook on other outputs." Similarly, Understandability is frequently mentioned by users, and they consider it the key to the effectiveness of XAI. For instance, P2 mentioned, "Originally, I have doubts about the outputs of AI, and I turn to XAI systems to seek answers. However, if it is still difficult to understand, then one would have to seek XAI for XAI."
   
\textbf{2. Sensitivity is the most important XUI design principle,} with a weight of 0.3296, but Flexibility also holds a weight of 0.3014. Sensitivity and Flexibility are crucial for users' trust and understandability attributes. This data is corroborated by responses to Q2 during the interviews. Users perceive Sensitivity and Flexibility as sources of subjective and objective trust, respectively. Users praise the user experience of Sensitivity because it increases the content of relevant information and makes the system feel intelligent. For instance, P4 mentioned, "The XUI design with sensitivity makes me feel very relaxed. I don't need to repeatedly self-report, and the information is mostly tailored to my specific situation, reducing a lot of unnecessary information." Flexibility is well-received because it eliminates ambiguity, P12 mentioned, "For illnesses, I have both resistance and anxiety. Flexibility can meet my needs well and eliminate many doubts I have about AI conclusions."

\begin{table}[h]
  \centering
  \renewcommand{\arraystretch}{1.4}
  \caption{the weights of users for five XAI user experience standard}
  \label{tab6}
  \begin{tabular}{c|c|c|c|c}
    \hline
    Satisfaction & Trust & Persuasiveness & Efficiency & Understandability \\
    \hline
    0.1604 & 0.2903 & 0.1663 & 0.1433 & 0.2398 \\
    \hline
  \end{tabular}
\end{table}

\begin{table}[h]
  \centering
  \renewcommand{\arraystretch}{1.4}
  \caption{the weights for different design principles of XUI}
  \label{tab7}
  \begin{tabular}{c|c|c|c}
    \hline
    Design principle A & Design principle B & Design principle C & Design principle D  \\
    \hline
    0.1549 & 0.2140 & 0.3014 & 0.3296 \\
    \hline
  \end{tabular}
\end{table}

\begin{table}[h]
  \centering
  \renewcommand{\arraystretch}{1.4}
  \caption{The weights of design principles under the five XAI user experience standards}
  \label{tab8}
  \begin{adjustbox}{width=\textwidth}
    \begin{tabular}{c|c|c|c|c}
      \hline
      & Design principle A & Design principle B  & Design principle C & Design principle D \\
      \hline
      Persuasiveness & 0.0991 & 0.2816 & 0.3209 & 0.2984 \\
      Satisfaction & 0.1528 & 0.1954 & 0.2722 & 0.3796 \\
      Trust & 0.1073 & 0.1955 & 0.3397 & 0.3576 \\
      Efficiency & 0.4135 & 0.2056 & 0.2763 & 0.1046 \\
      Understandability & 0.0844 & 0.1603 & 0.3609 & 0.3944 \\
      \hline
    \end{tabular}  
  \end{adjustbox}
\end{table}

\section{Discussion}

In this section, we will provide further insights and discussions based on the experimental results, summarizing our experiences and offering valuable information for XUI design. We will discuss on this in three subpoints :

\textbf{1, Users demand "correct" information.} Users are more concerned about whether the information provided is "correct" (meeting their specific needs) rather than just being comprehensive or persuasive. The weight of the Sensitivity principle, which provides context-sensitive responses, is 0.3796 under the satisfaction criterion, compared to the Flexibility principle, which provides more detailed information with a weight of 0.2722. This trend is also observed under the trust criterion. Users tend to prefer information that aligns with their specific needs rather than an abundance of information. This aligns with previous research findings that users dislike explanatory forms requiring more effort.
 
\textbf{2, User experience is the core of XAI applications.} Analysis of the overall weights for the four XUI designs reveals that designs centered around user-centric principles (such as Design principles C and D) often outperform designs less focused on user experience (such as Design principle A, which is more algorithm-centric). This further emphasizes the importance of HCXAI, suggesting that XAI development should prioritize user needs. If detached from user requirements, XAI may lose its practical value.

\textbf{3, Diverse Demands.} In our research, we discovered that differentiated needs are a highly significant issue, primarily classified into two types:\paragraph{Individual Differentiation:}Almost every individual exhibits different preferences. XUI outputs should emphasize differentiation. For instance, Participants P7 and P15 prefer the "Naturalness" design principle, unlike others. They believe that adding other forms is a waste of time when textual descriptions are correct.\paragraph{Scenario Differentiation:}The data results indicate significant fluctuations in the weights of the four principles under different XAI user experience criteria. For example, "Naturalness" performs relatively poorly under other user experience criteria but excels under the efficiency criterion. Therefore, flexible application of different XUI design principles is recommended based on diverse scenarios and requirements.

\section{Limitation and Future Work}
In the following section, we will discuss the limitations of our study and potential directions for future research :

\textbf{1. Lack of consideration for the combination of design principles.} The four principles discussed in this study are entirely combinable, yet our research treats them as independent principles to explore their individual importance. While there are challenges related to the limited UI design space, future research could investigate the impact of combining multiple XUI explanation principles on users, for a more precise response to user needs.

\textbf{2. Limited consideration of scalability.} Due to constraints in the experimental environment and controlled variables, our study has limitations in terms of scalability. Firstly, it only focuses on conversational AI interfaces, neglecting exploration into other forms of AI interfaces such as XR interfaces or natural interfaces. Secondly, the study does not account for changes over extended usage periods. In high-frequency usage scenarios, user demands may change, and the weight of factors like "efficiency" could correspondingly increase. Lastly, the study has a single-use scenario, real-life situations are more complex, with diverse user needs across different usage scenarios. Future research could explore the scalability of various XUI design principles in more detail.

\section{Conclusion}
In this paper, we conducted a study on the weighting of XUI design principles and summarized lightweight XAI user experience standards for non-expert users. Our contributions include providing weighted rankings for design principles aimed at enhancing the XUI user experience and offering guidance for practitioners in allocating XUI design space reasonably. Additionally, we provided a lightweight summary of XAI user experience standards for non-expert users from the perspective of HCXAI, serving as a reference for future researchers. As the widespread use of LLMs continues, the demand for XAI is expected to grow, especially among non-expert users. Our study provides valuable insights for specific XUI designs and contributes to improving the user experience of XAI through UI design. In the future, we hope these research findings will guide XUI design and encourage more researchers to engage in user experience studies in the field of XAI.

\section{Acknowledgments}

\begin{flushleft}
This version of the contribution has been accepted for publication, after peer review but is not the Version of Record and does not reflect post-acceptance improvements, or any corrections. Use of this Accepted Version is subject to the publisher’s Accepted Manuscript terms of use: \url{https://www.springernature.com/gp/open-research/policies/accepted-manuscript-terms}.
\end{flushleft}

%
%

%
%
%
%

\end{document}